\documentclass{pnastwo}
\usepackage[dvips]{graphicx}
\usepackage{amsmath}
\usepackage{amsfonts}
\usepackage{amssymb}
\usepackage{wasysym}

\def\be{\begin{equation}}
\def\ee{\end{equation}}
\begin{document}

\conflictofinterest{The authors declare no conflict of interest}

\title{Learning as a phenomenon occurring in a critical state}

\author{Lucilla de Arcangelis
\thanks{To whom correspondence should be addressed. Email: <dearcangelis@na.infn.it>}
\affil{1} 
{Institute Computational Physics for Engineering Materials,
ETH, Schafmattstr.6, 8093 Z\"urich, CH}
\affil{2}
{Department of Information Engineering and CNISM,
Second University of Naples, 81031 Aversa (CE), Italy}
and Hans J. Herrmann\affil{1}{}}

\maketitle
\begin{article}
\begin{abstract}
Recent physiological measurements have provided clear evidence about scale-free
avalanche brain activity and EEG spectra, feeding the classical enigma of how 
such a chaotic system can ever learn or respond in a controlled and reproducible way.
Models for learning, like neural networks or perceptrons, have traditionally avoided strong 
fluctuations. Conversely, we propose that brain activity having features typical of 
systems at a critical point, represents a crucial ingredient for learning.
We present here a study which provides novel insights toward the understanding
of the problem.
Our model is able to reproduce quantitatively the experimentally observed 
critical state of the brain and, at the same time, learns and remembers logical rules
including the exclusive OR (XOR), which has posed difficulties to several previous attempts.
We implement the model on a network with topological properties close to the 
functionality network in real brains. 
Learning occurs via plastic adaptation of synaptic strengths 
and exhibits universal features.
We find that the learning performance and the average time required to learn  
are controlled by the strength of plastic adaptation, in a way independent 
of the specific task assigned to the system.
Even complex rules can be learned provided that the plastic adaptation is sufficiently slow.
\end{abstract}

\keywords{learning | scale-free network | self-organized criticality | scaling}

Spontaneous activity is an important property of the cerebral cortex 
that can have a crucial role in information processing and storage.
Recently it has been shown that a novel spatio-temporal form of spontaneous activity is 
neuronal avalanches, which can involve from a few to a very large number of neurons. 
These bursts of firing neurons have been first observed \cite{beg1,beg2} in organotypic
cultures from coronal slices of rat cortex, where the size and duration of
neuronal avalanches follow power law distributions with very stable exponents.
The presence of a power law behaviour is the typical feature of a system acting in a 
critical state \cite{sta}, where large fluctuations are present and the response does
not have a characteristic size.
The same critical behaviour, namely the same power law exponents,
has been recently measured also {\it in vivo} 
from superficial layers of cortex in anesthetized rats
during early 
post-natal development \cite{pnas}, and awake adult rhesus monkeys \cite{pnas2}, 
using micro-electrode array recordings. Results confirm that indeed spontaneous cortical 
activity adjusts in a critical state where the spatio-temporal organization of avalanches 
is scale invariant. Moreover, the investigation on the spontaneous activity
of dissociated neurons from different networks as rat hippocampal neurons \cite{maz},
rat embryos \cite{pas} or leech ganglia \cite{maz}, has also confirmed the robustness of
this scaling behaviour. 
In all these cases, the emergence of power law distributions has been interpreted 
in terms of self-organized criticality (SOC) \cite{bak}.
The term SOC usually refers to a mechanism of
slow energy accumulation and fast energy redistribution driving the system
toward a critical state, where the avalanche extensions and durations
do not have a characteristic size.

\begin{figure}
\includegraphics[width=8cm]{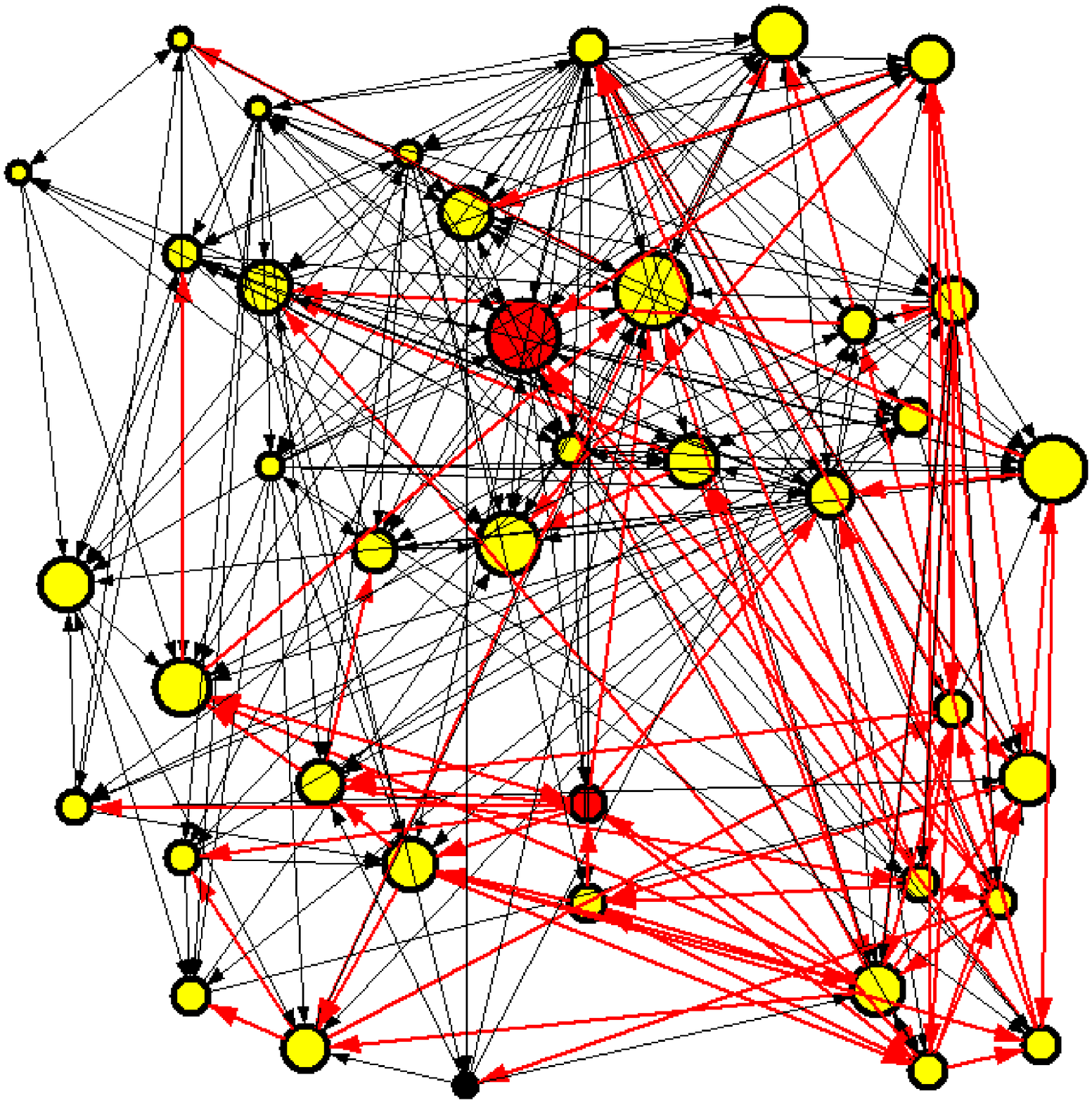}
\caption{ 
One neuronal avalanche in the scale-free network.
40 neurons are connected by directed bonds (direction indicated by 
the arrow at one edge), representing the synapses. 
The size of each neuron is proportional to the number of in-connections, 
namely the number 
of dendrites. The two red neurons are the two input sites, whereas the black
neuron is the output. Connections involved in the avalanche propagation are
shown in red, whereas inactive connections are black.
}
 \label{fig1}
\end{figure}

The understanding of the fundamental relations between electro-physiological
activity and brain organization with respect to performing even simple tasks,  
is a long-standing fascinating question. A number of theoretical
models \cite{ami}
have been proposed for learning, from the simple perceptron \cite{ros} to Attractor Neural 
Networks (ANNs) \cite{hop} of artificial two-state neurons \cite{mcc}. In these models
the state of the ``brain" is the snapshot of the ensemble of the individual states of 
all neurons, which 
explores phase space following an appropriate dynamics and eventually recovers memories.
The ability of the brain to self-organize 
connections in an efficient way is a crucial ingredient in biologically plausible
models. The breakthrough of Hebbian plasticity, postulating synapse 
strengthening for correlated activity at the pre- and post-synaptic neuron and
synapse weakening for decorrelated activity, triggered 
the development of algorithms for neuronal learning and memory, as, for instance, 
``reinforcement learning"
\cite{bar} or error back-propagation \cite{rum}, leading for the first time to the XOR rule
learning. Recent results have shown that extremal dynamics, where
only the neuron with the largest input fires, and uniform negative feedback are sufficient 
ingredients to learn the following task: to identify the right connection between 
an input and an output node \cite{bakchi1,bakchi2}. Similarly, low activity
probabilistic firing rules, where again a single neuron fires at each step of the 
iteration, together with a uniform negative feedback plastic adaptation acting on time 
scales slower than the neuron
firing time scale, enables learning the XOR rule without error back-propagation \cite{kle}.
Both results suggest that the system learns by mistakes, namely depression rather than 
enhancement of synaptic strength is the crucial mechanism for learning.
However, in both studies a single neuron fires at each 
step of the evolution, not in complete agreement with recent experimental discoveries.
Cooperative effects leading to self-organization and learning are completely neglected 
in the aforementioned approaches.

Operating at a critical level, far from an uncorrelated subcritical or
a too correlated supercritical regime, may optimize information management and
transmission in real brains \cite{tur,beg1,kin,eur}, as recently confirmed by
experiments \cite{she}. 
Moreover, a recent study of visual perceptual
learning has evidenced that training to a specific task
induces dynamic changes in the functional connectivity
able to ``sculpt" the spontaneous activity of the resting human brain and
to act as a form of ``system memory" \cite{lew}. It is therefore
tempting to investigate  the role that critical behaviour plays in
the most important task of neuronal networks, namely learning and memory.
The emergence of a critical state with the same critical behaviour found experimentally
has been recently reproduced by a neuronal network model
based on SOC ideas \cite{br1,br2}. The model implements several physiological properties
of real neurons: a continuous membrane potential, firing at threshold, synaptic plasticity
and pruning. Extensive numerical studies on regular, small world and scale free networks
have shown that indeed the system exhibits a robust critical behaviour. The distributions
of avalanche size and duration scale with exponents independent of model parameters and
in excellent agreement with experimental data (Fig.2). 
More precisely, the distribution of avalanche sizes, measured experimentally either
in terms of number of active electrodes or summed local field potentials in a 
micro-electrode array \cite{beg1,beg2}, decreases with an exponent $-1.5$,
whereas the distribution of avalanche temporal durations decreases with an exponent close
to  $-2.0$. A critical avalanche activity has been also found on  fully connected
\cite{lev} and random networks \cite{ter}. Moreover, the temporal
signal for electrical activity and  the power spectrum of the resulting
time series have been compared  with EEG data \cite{br1,br2}.
The spectrum exhibits a power law behaviour, $P(f)\sim f^{-0.8}$, 
with an exponent in good agreement with EEG medical data \cite{nov} and physiological
signal spectra for other brain controlled activities \cite{hau}.
This model therefore seems to capture many of the essential ingredients of spontaneous
activity, as measured in cortical networks.

Here we study the learning performance of a neuronal network acting in a critical state.
The response of the system to external stimuli is therefore scale-free, i.e. no 
characteristic size in the number of firing neurons exists. 
The approach reproduces closely the physiological mechanisms of neuronal
behaviour and is implemented on a plausible network having topological properties 
similar to the brain functionality network. Neuronal activity is a collective
process where all neurons at threshold can fire and self-organize an efficient path
for information transmission. Plastic adaptation is introduced via a non uniform negative 
feedback procedure with no error back propagation. 

\section{The model}

We consider $N$ neurons positioned at random in a two dimensional
space. Each neuron is characterized by the potential $v_i$.
Connections among neurons are established by assigning  
to each neuron $i$ a random out-going connectivity degree, $k_{out_i}$. 
The distribution of the number of out-connections is then chosen in agreement 
with the experimentally measured 
properties of the functionality network \cite{chia2} in human adults.
Functional magnetic resonance imaging has indeed shown that this 
network has universal scale free properties, namely it exhibits a scaling behaviour
$n(k_{out})\propto k_{out}^{-2}$, independent of the different tasks performed by 
the patient. We adopt this distribution for
the number of pre-synaptic terminals of each neuron, over the range of possible values 
between $k^{min}_{out}$ and $k^{max}_{out}=100$, as in experimental data.
Two neurons are then connected according to a distance dependent probability,
$p(r)\propto e^{-r/r_0}$, where $r$ is their spatial distance \cite{roe} and
$r_0$ a typical edge length.
To each synaptic connection we then assign an
initial random strength $g_{ij}$, where $g_{ij}\neq g_{ji}$, and an
excitatory or inhibitory character, with a fraction $p_{in}$ of inhibitory synapses.
An example of such a network is shown in Fig.1.

The firing dynamics implies that, whenever at time $t$ the
value of the potential at a site $i$ is above a certain threshold
$v_i \geq v_{\rm max}=6.0$,
approximately equal to $-55mV$ for real neurons, the neuron sends action potentials
leading to the production of an amount
of neurotransmitter proportional to $v_i$. As a consequence, the
total charge released by a neuron is proportional to the number
of synaptic connections, $q_i\propto v_i k_{out_i}$.
Each connected neuron receives charge in proportion to the strength of the synapse $g_{ij}$
\be
v_j(t+1)=v_j(t)\pm \frac{q_i(t)}{k_{in_j}}\frac{g_{ij}(t)}{\sum_k g_{ik}(t)} \label{prop}
\ee
where $k_{in_j}$ is the in-degree of neuron $j$ and the sum is extended to all out-going
connections of $i$. 
In Eq.(1) it is assumed that the received charge is distributed
over the surface of the soma of the post-synaptic neuron, proportional
to the number of  in-going terminals $k_{in_j}$. The plus or minus sign in Eq.(1)
is for excitatory or inhibitory synapses, respectively.
After firing a neuron is set to a zero resting potential and
in a refractory state lasting $t_{ref}=1$ time step, during which it is unable to receive
or transmit any charge. We wish to stress that the unit time step in Eq.(1) does not 
correspond to a real time scale, it is simply the time unit for charge propagation 
from one neuron to the connected ones. 
In real system  this time could vary and be as large as 100 ms 
for longer firing periods.
The synaptic strengths have initially a random value $g_{ij}\in [0.5,1.0]$, whereas
the neuron potentials are uniformly distributed random numbers between
$v_{\rm max} - 1$ and $v_{\rm max}$. 
Moreover, a small random fraction ($10\%$) of neurons is chosen to be boundary sites,
with a potential fixed to zero, playing the role of sinks for the charge.

\begin{figure}
\vskip+0.2cm
\includegraphics[width=8cm]{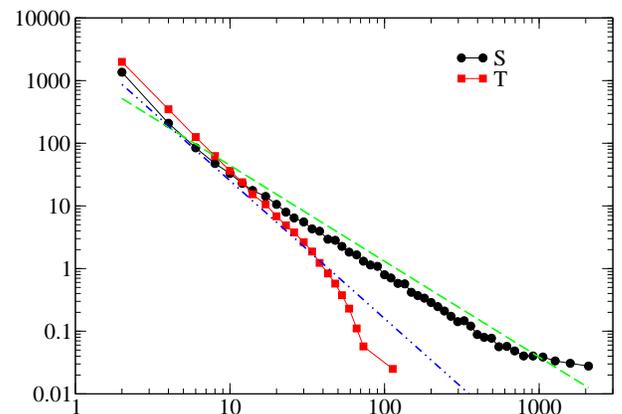}
\caption{
Demonstration of the critical behaviour of the neuronal network avalanche activity.
The distribution of the sizes of the avalanches of firing neurons, $n(S)$ (circles), 
follows a power law behaviour with an exponent $1.5\pm 0.1$ (dashed line). 
The size is measured as
the number of firing neurons. The distribution of avalanche  durations, $n(T)$ (squares),
exhibits a power law behaviour with an exponent $2.2\pm 0.2$ (dot-dashed line), 
followed by an exponential cutoff. Data are obtained for 40 realizations of a network 
of 4000 neurons with $p_{in}=0.05$.
}
 \label{fig2}
\end{figure}

In order to start activity we identify input neurons at which the imposed signal  
is applied and the output neuron at which the response is monitored. 
These nodes are randomly placed inside the network under the condition that they are not 
boundary sites and they are mutually separated on the network by $k_d$ nodes. $k_d$ represents the 
chemical distance on the network and plays the role of the number of hidden layers 
in a perceptron. We test the ability of the network to
learn different rules: AND, OR, XOR and a random rule RAN which associates to all possible
combinations of binary states at three inputs a random binary output. More precisely,
the AND, OR and XOR rules are made of three input-output relations (we disregards the double 
zero input which is a trivial test leading to zero output), whereas the RAN rule 
with three input sites implies a sequence of seven input-output relations. A single learning 
step requires the application of the entire sequence of states at 
the input neurons, monitoring the state of the output neuron. 
For each rule the binary value 1 is identified with the output neuron firing, namely
the neuron membrane potential at a value greater or equal to $v_{\rm max}$ at some 
time during the activity.
Conversely, the binary state 0 at the output neuron corresponds to the
physiological state of a real neuron which has been depolarized by incoming ions but
fails to reach the firing threshold membrane potential during the entire avalanche 
propagation. 
Once  the input sites are stimulated, their activity may bring to threshold other
neurons and therefore lead to avalanches of firings. We impose no restriction on the number 
of firing neurons in the propagation and
let the avalanche evolve to its end according to Eq.(\ref{prop}). 
If at the end of the avalanche the propagation of charge did not reach the 
output neuron,
we consider that the state of the system was unable to respond to the given stimulus,
and as a consequence to learn. We
therefore increase uniformly the potential of all neurons by units of a small quantity,
$\beta = 0.01$, until the configuration reaches a state where the output neuron is first
perturbed.
We then compare the state of the output neuron with the desired output. 
Namely we follow the evolution in phase space of the initial state of the system
and verify if the non-ergodic dynamics has led to an attractor associated with the 
right answer.

\begin{figure}
\vskip+0.2cm
\includegraphics[width=8cm]{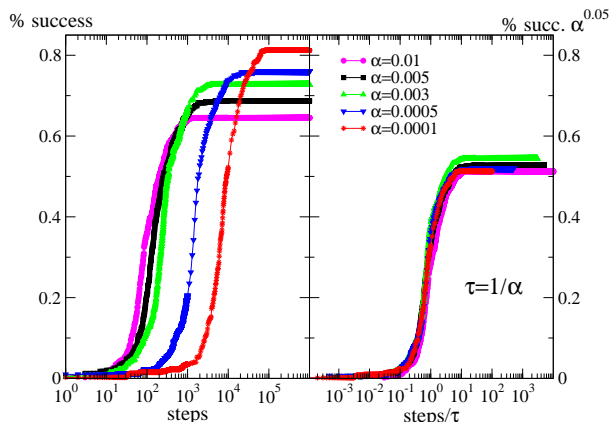}
\caption{ 
Demonstration of the learning ability of the neuronal network.
{\it (Left)} Percentage of configurations learning the XOR rule
as function of the number of learning steps for different plastic adaptation strengths
$\alpha$ (decreasing from bottom to top). 
Data are for 400 realizations of a network with $N=1000$ neurons,
$p_{in}=0.1$, $r_0=15$, $k_{min}=3$ and $k_d=5$.
{\it (Right)} Collapse of the curves by rescaling the number of learning steps 
by the characteristic
learning time $\tau=1/\alpha$ and the percentage of success by $\alpha ^{-0.05}$.
}
 \label{fig3}
\end{figure}

\subsection{Plastic adaptation}

Plastic adaptation is applied to the system according to a non uniform 
negative feedback algorithm.
Namely, if the output neuron is in the correct state according to the rule, 
we keep the value of synaptic strengths. Conversely, if the response is wrong we 
modify the strengths of those synapses involved in the information propagation
by $\pm \alpha/d_k$, where $d_k$ is the chemical distance of the presynaptic neuron
from the output neuron. Here $\alpha$ represents the ensemble of
all possible physiological factors influencing synaptic plasticity.
The sign of the adjustment depends on the mistake made by the system:
If the output neuron fails to be in a firing state we increase the used synapses by
a small additive quantity proportional to $\alpha$. Synaptic strengths are instead 
decreased by if the expected output 0 is not fulfilled. 
Once the strength of a synapse is below an assigned small value
$g_t=10^{-4}$, we remove it, i.e. set its strength equal to zero, which corresponds to
the so-called {\it pruning}. This ingredient is very important as
since decades the crucial role
of selective weakening and elimination of unneeded connections in adult  learning  has 
been recognized \cite{you,cha}.
The synapses involved in the signal propagation and responsible for the wrong
answer, are therefore not adapted uniformly but inversely proportional 
to the chemical distance from
the output site. Namely, synapses directly connected to the output neuron receive the
strongest adaptation $\pm \alpha$. 
This adaptation rule
intends to mimic the feedback to the wrong answer triggered locally at the
output site and propagating backward towards the input sites.
This could be the case, for instance, of some hormones strongly interfering with learning and
memory, as dopamine suppressing LTD \cite{otm} or adrenal hormones enhancing LTD \cite{cou}.
Moreover a new class of messenger molecules as nitric oxide has been found to have an
important role in plastic adaptation \cite{rey}. For all these agents, released at the 
output neuron, the concentration is reduced with the distance from the origin.
In our neuronal network simulation this non uniform adaptation has a crucial role
since it prevents, in case of successive wrong positive answers,
synapses directly connected to the input sites to decrease excessively, hindering any
further signal transmission. This plastic adaptation is a non-Hebbian form 
of plasticity and can be interpreted as a subtractive form of 
synaptic scaling \cite{abb}, where synapses are changed by an amount independent of 
their strength. The procedure mimics the
performance of a {\it good critic} who does not tell the system which neurons should 
have fired or not. However it tells more than just ``right" or ``wrong",
it expresses an evaluation on the type of error. 
Finally, we wish to stress that this model naturally sets the system in a critical state
and therefore the study of the response of the system in a subcritical or
supercritical state requires the introduction of additional parameters.
We can however suppose that in both cases learning becomes a more difficult
task. For instance, in a subcritical state, being the size of neuronal
avalanches smaller, it would be more complex to generate a firing state in
the output site. Conversely, in a supercritical state it would be more difficult to
generate a non-firing state in the output site.

\begin{figure}
\vskip+0.2cm
\includegraphics[width=8cm]{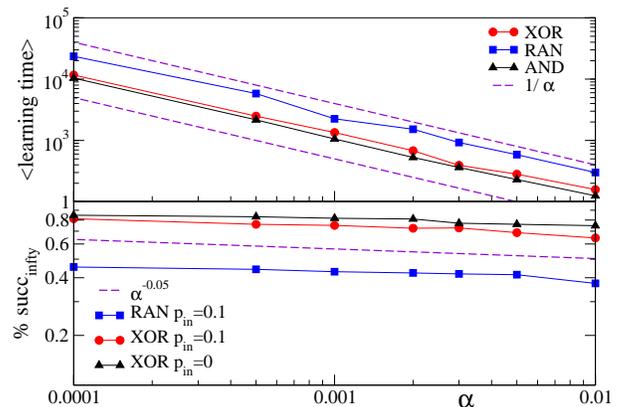}
\caption{ 
Demonstration of the scaling behaviour of learning time and success.
{\it (Top)} Scaling of the average learning time for different rules
as function of $\alpha$. 
{\it (Bottom)} Scaling of  the asymptotic percentage of configurations 
learning different rules as function of $\alpha$. Data are obtained for 
$k_{min}=3$, $k_d=5$ and $p_{in}=0.1$. Both quantities follow a power law 
with an exponent independent of 
the rule and the percentage of inhibitory synapses.
}
\label{fig4}
\end{figure}

\section{Results and Discussion}

We analyse the ability of the system to learn the different rules. Fig. 3 shows the 
fraction of configurations learning the XOR rule versus the number of learning steps
 for different
values of the plastic adaptation strength $\alpha$. We notice that the larger the value of 
$\alpha$ the sooner the system starts to learn the rule, however the final percentage of 
learning configurations is lower. The final rate of success increases as the strength
of plastic adaptation decreases. This result is due to 
the highly non linear dynamics of the model, where firing activity is an all or none 
event controlled by the threshold. Very slow plastic adaptation allows then to tune finely
the role of the neurons involved in the propagation and eventually recover the right answer.
Moreover, very slow plastic adaptation also makes the system more stable with respect
to noise, since too strong synaptic changes  may perturb excessively the evolution 
hampering the recovery of the right answer.   
The dependence of the learning success on the plasticity strength is found consistently 
for different values of the parameters $k_d$, $k_{min}$ and $p_{in}$, where a higher
percentage of success is observed in systems with no inhibitory synapses. 
Moreover, the dependence on the plastic adaptation $\alpha$ is a 
common feature of all tested rules. Data indicate that the easiest rule to learn is OR, 
where a 100\% percentage of success can be obtained. AND and XOR present similar 
difficulties and lead both to a percentage of final success around 80\%, whereas the most 
difficult rule to learn is the RAN rule with three inputs where only 50\% of final success 
is obtained. This different performance is mainly due to the higher number of inputs, 
since the system
has to organized a more complex path of connections leading to the output site.

The most striking result is that all rules give a higher percentage of success for
weaker plastic adaptation. 
Indeed this result is in agreement with recent experimental findings on visual
perceptual learning, where better performances are measured when minimal changes in
the functional network occur as a result of learning \cite{lew}. 
We characterize the learning ability of a system 
for different rules by the average learning time, i.e. the average number of times a 
rule must be applied to obtain the right answer, and the asymptotic percentage of 
learning configurations. This is determined as the percentage of learning configurations at
the end of the teaching routine, namely after $10^6$ applications of the rule. 
Fig. 4 shows that the average learning time scales as
$\tau\propto 1/\alpha$ for all rules and independently of parameter values. 
Since some configurations never learn and do not contribute to the average learning
time, we also evaluate the median learning time
which exhibits the same scaling behaviour as the average learning time. 
The asymptotic percentage of success increases by decreasing $\alpha$  
as a very slow power law, $\propto \alpha^{-0.05}$. Since this quantity has
a finite upper bound equal to unity, this scaling suggests that in a finite, even if 
very long,
time any configuration could learn the rule by applying an extremely slow plastic 
adaptation.
It is interesting to notice that a larger fraction of systems with no
inhibitory synapses finds the right
answer and the average learning time for these systems is slightly shorter. 
We understand this result by considering that for only excitatory synapses 
the system more easily selects a path of strong enough synapses connecting
inputs and output sites and giving the right answer. 
Conversely, the presence of inhibitory synapses may
lead to frustration in the system as not all local interactions contribute in
the optimal way to provide the right answer and the system has to find
alternative paths.
We check this scaling behaviour by appropriately rescaling the axes in Fig. 3. The curves 
corresponding to different $\alpha$ values indeed all collapse onto a unique scaling 
function. 
Similar collapse is observed for the OR, AND and RAN rules and for different
parameters $k_d$, $k_{min}$ and $p_{in}$. In fact, two different cases of
distributions of inhibitory synapses, one in which they are chosen randomly
among all synapses, the other where certain randomly chosen neurons have all
outgoing synapses inhibitory, provide equivalent results.
The learning dynamics shows therefore
universal properties, independent of the details of the system or the specific task assigned.

The learning behaviour 
is sensitive to the number of neurons involved in the propagation of
the signal, and therefore depends on the distance between input and output
neurons and the level of connectivity in the system.
We then investigate the effect of the parameters $k_d$ and $k_{min}$ on the
performance of the system.
Fig. 5 shows the percentage of configurations learning the XOR rule for
different minimum values of the neuron out degree. Systems with larger $k_{min}$ have a
larger
average number of synapses per neuron, producing a more branched network. 
The presence of several alternative paths facilitates information 
transmission from the 
inputs to the output site. However, the participation of more branched
synaptic paths 
in the learning process may delay the time the system first gives the 
right answer. As expected the performance of the system improves as the minimum 
out-connectivity degree increases, with the asymptotic percentage of success scaling as
$\sim k_{min}^{0.4}$. The dependence of the learning performance on the level
of connectivity is confirmed by the analysis of systems with different number
of neurons $N$, the same out-degree distribution and the same set of 
parameters. We verify that larger systems exhibit better performances. 
In larger systems, in fact, the number of hubs, i.e. highly
connected neurons, increases improving the overall level of connectivity.
Indeed, the existence of complex patterns of 
activation has been recently
recognized as very important in linking together large scale networks in visual 
perceptual learning\cite{lew}.

\begin{figure}
\vskip+1.5cm
\includegraphics[width=8cm]{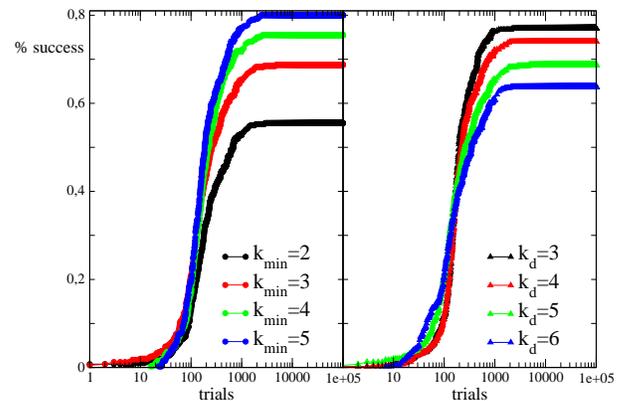}
\caption{ 
Demonstration of the dependence of the learning performance on the minimum 
connectivity degree 
and the chemical distance between input and output sites.
{\it (Left)}  Percentage of configurations learning the XOR rule
as function of the number of learning steps,
for $\alpha=0.005$ in systems with $k_d=5$ and different values of $k_{min}$
(increasing from bottom to top).
{\it (Right)} Percentage of configurations learning the XOR rule
as function of the number of learning steps,
for $\alpha=0.005$ in systems with $k_{min}=3$ and different values of $k_d$
(decreasing from bottom to top).
Data are for 400 realizations of a network with $N=1000$ neurons,
$p_{in}=0.1$. 
}
\label{fig5}
\end{figure}

On the other hand, also the chemical distance between the input and output
sites has a very important role, as the number of hidden layers in a perceptron.
Indeed, as $k_d$ becomes larger (Fig. 5), the length of each branch in a path involved 
in the learning process increases. 
As a consequence, the system needs a higher number of tests to first give the 
right answer and a lower fraction of configurations learns the rule after the same 
number 
of steps. The percentage of learning configurations after $10^6$ applications is found,
as expected, 
to decrease as $\sim k_d^{-0.3}$ and similar behaviour is detected for the OR, AND and 
RAN rules.

\subsection{Learning stability and Memory}

The existence of systems that are unable to learn, even after many learning steps, raises
intriguing questions about the learning dynamics. We question what happens
when a second chance is given 
to the configurations failing the right answer. We then restart the learning routine 
after 
imposing a small change in the initial configuration of voltages. This small perturbation
leads to about 25\% more configurations 
learning the rule. The initial state of the system can therefore influence the ability 
to learn, especially for complex rules as XOR or RAN. On the other hand, the analysis
of the out-degree distribution in configurations which did and did 
not give the right answer indicates that ``dumb" configurations  tend to have less highly 
connected nodes  than the ```smart" ones. Namely, giving repeatedly wrong positive
answers leads to 
pruning of several synapses. This affects in particular the  highly connected neurons, 
which have a crucial role in identifying the right synaptic learning path.
Finally we test the ability of the configurations that do learn to remember the 
right 
answer once the initial configuration is changed. The memory performance 
of the system is 
expected to depend on the intensity of the variation imposed, namely on the extension of
the basin of the attraction of states leading to the right answer. The system is able to 
recover the right answer in more than 50\% of the configurations if a very small 
perturbation
(of the order of $10^{-3}$) is applied to all neurons or else a larger one (of the order 
of $10^{-2}$) to 10\% of neurons. The system has a different memory ability depending 
on the rule: almost all configurations remember OR, whereas typically 80\% remember AND 
and at most 70\% the XOR rule.  

\section{Conclusions}

In conclusion, we investigate the learning ability of a model able to reproduce 
the critical avalanche activity as observed for spontaneous activity in 
{\it in vitro} and {\it in vivo} cortical networks. The ingredients of 
the model are close to most functional and topological properties of real neuronal 
networks. The implemented learning dynamics 
is a cooperative mechanism where all neurons contribute to select the right answer
and negative feedback is provided in a non-uniform way. Despite the complexity of
the model and the high number of degrees of freedom involved at each step of the
iteration, the system can learn successfully even complex rules as XOR or a random 
rule with three inputs. 
In fact, since the system acts in a critical state, the response to a given input can be
highly flexible, adapting more easily to different rules.
The analysis of the dependence of the performance of the
system on the average connectivity  confirms that 
learning is a truly collective process, where a high number of neurons may be involved
and the system learns more efficiently if more branched paths are possible.
The role of the plastic adaptation 
strength, considered as a constant parameter in most studies, provides a striking new
result: 
The neuronal network has a ``universal" learning dynamics, even complex rules 
can be learned provided that the plastic adaptation is sufficiently slow.
This important requirement for plastic adaptation is confirmed by recent experimental 
results \cite{lew} showing that the learning 
performance, in humans trained to a specific visual task, improves when minimal 
changes occur in the functionality network. Stronger modifications of the network do not
necessarily lead to better results.

\end{article}
\end{document}